\documentclass[final,5p,times,twocolumn]{elsarticle}

\usepackage{amssymb}
\usepackage{float}
\usepackage{booktabs}
\usepackage{amsmath,amstext,amssymb}
\usepackage{subfigure}
\usepackage{natbib}

\journal{Physics Letters A}

\begin{document}

\begin{frontmatter}

%% Title, authors and addresses

%% use the tnoteref command within \title for footnotes;
%% use the tnotetext command for the associated footnote;
%% use the fnref command within \author or \address for footnotes;
%% use the fntext command for the associated footnote;
%% use the corref command within \author for corresponding author footnotes;
%% use the cortext command for the associated footnote;
%% use the ead command for the email address,
%% and the form \ead[url] for the home page:
%%
%% \title{Title\tnoteref{label1}}
%% \tnotetext[label1]{}
%% \author{Name\corref{cor1}\fnref{label2}}
%% \ead{email address}
%% \ead[url]{home page}
%% \fntext[label2]{}
%% \cortext[cor1]{}
%% \address{Address\fnref{label3}}
%% \fntext[label3]{}

\title{Characteristic times for the Fermi-Ulam Model}

%% use optional labels to link authors explicitly to addresses:
%% \author[label1,label2]{<author name>}
%% \address[label1]{<address>}
%% \address[label2]{<address>}

\author{Joelson D. Veloso Hermes$^{\rm 1,}$$^{\rm 2}$}
\ead{joelson.hermes@ifsuldeminas.edu.br, Phone:+55 35 3464 1200}

\author{Edson D.\ Leonel$^{\rm 2}$}
%\ead{edleonel@rc.unesp.br}
\address{$^{\rm 1}$Instituto Federal de Educa\c c\~ao, Ci\^encia e Tecnologia 
do Sul de Minas Gerais - IFSULDEMINAS, Inconfidentes - Brazil\\
$^{\rm 2}$Departamento de F\'isica, UNESP - Univ. Estadual Paulista, Av.24A,
1515 - Bela Vista - 13506-900 - Rio Claro - SP - Brazil\\}

%\pacs{05.45.Pq, 05.45.Tp}

\begin{abstract}
The mean Poincar\'e recurrence time as well as the Lyapunov time are measured 
for the Fermi-Ulam model. We confirm the mean recurrence time is dependent on 
the size of the window chosen in the phase space to where particles are allowed 
to recur. The fractal dimension of the region is determined by the slope of the 
recurrence time against the size of the window and two numerical values were 
measured: (i) $\mu=1$ confirming normal diffusion for chaotic regions far from 
periodic domains and; (ii) $\mu=2$ leading to anomalous diffusion measured 
near periodic regions, a signature of local trapping of an ensemble of 
particles. The Lyapunov time is measured over different domains in the phase 
space through a direct determination of the Lyapunov exponent, indeed being 
defined as its inverse.
\end{abstract}


\end{frontmatter}

\date{today}

%%
%% Start line numbering here if you want
%%
%\linenumbers

%% main text

\section{Introduction}
\label{sec1}

Diffusion of particles has intrigued scientists of different areas over the 
years. Applications are the wider as possible ranging from medicine \cite{p2} 
where a specific medical or chemical drug diffuses in blood to reach its 
destiny, water infiltration \cite{p5} in the surface of the planet caring 
chemical elements from pesticides to the water table, pollen diffusion of 
plants \cite{p3}, pollution in air \cite{p4} or in water \cite{b1} and many 
others. In dynamical systems diffusion can be treated as via the solution of the 
diffusion equation \cite{b2} leading to results proving scaling invariance in 
chaotic systems \cite{celia}. Moreover, diffusion in Hamiltonian systems is 
also connected to Poincar\'e recurrence \cite{lichtenberg1992} which is defined 
as the time a particle spends moving along the phase space to return to a 
specific region to where it has passed earlier. It is known \cite{Harsoula2018a} 
that such a time obeys specific laws that confirm the existence of stickiness 
\cite{altmann1} and hence anomalous diffusion \cite{buni1} or normal diffusion 
\cite{zas1}. 

Whenever observing stickiness, chaotic dynamics is also present. One of the 
basic tools to measure chaotic properties in nonlinear systems is the Lyapunov 
exponent $\lambda$ \cite{chaos_b}. It is based on the average separation in 
time of two nearby initial conditions. A characteristic time associated with it 
\cite{contopoulous_2} is the Lyapunov time $t_L=1/\lambda$. Moreover in 
Hamiltonian chaos the Liovuille's theorem \cite{liouville} warrants area 
preservation in the phase space. In mixed systems where chaos coexists with 
periodic islands and invariant tori in the phase space, a particle in the 
chaotic domain can not cross through the invariant tori nor get into the 
islands. This implies that once in the chaos, always in the chaos. It also 
yields in to an important property that a given particle may recur to a certain 
region in the phase space and that the time it spends to return to a specific 
domain is called as Poincar\'e recurrence time $t_r$. This characteristic time 
depends on the size of the region and on the type of the dynamics nearby it. 
The slope of the curve given by $t_r$ plotted against the size of the region 
gives the fractal dimension of the set of points of such a region and marks the 
kind of diffusion measured. In this paper we revisit the Fermi-Ulam model 
\cite{fermi} and we are seeking to understand and describe the behavior of the 
two characteristic times mentioned above, namely the Lyapunov and the 
Poincar\'e recurrence times. 

The model is composed of a particle confined to move inside of two rigid walls 
where one of them is fixed while the other one moves periodically in time. 
Collisions are assumed to be elastic in the sense that there is no lose of 
energy upon the impacts, hence the area of the phase space is preserved. When 
the particle has very low energy\footnote{We mention as very low when the 
energy of the particle is comparable with the energy of the moving wall, 
i.e., the velocity of the particle has the same magnitude of the maximum 
velocity of the moving wall.} the elapsed time between impacts is large. Hence 
there is no correlation between the phase of the moving wall at the impact $n$ 
as compared to the phase at the impact $(n+1)$. The absence of correlation 
between phases leads the velocity of the particle to grow. For the low energy 
regime a particle exhibits chaotic dynamics while with the growth of the 
velocity correlations between phases appear producing regularity in the phase 
space where islands of stability as well as invariant tori are observed. The 
lowest energy invariant spanning curve has crucial importance in limiting the 
size of the chaotic sea preventing the unlimited diffusion of the chaotic 
dynamics. This unlimited diffusion was believed to be observed in dynamical 
systems produced by collisions of an ensemble of particles with moving periodic 
boundary leading to a phenomena called as Fermi acceleration. The existence of 
the invariant spanning curve prevent this unlimited growth. It also imposes an 
interesting scaling invariance of the chaotic sea \cite{leonel} near a 
transition from integrability to non integrability \cite{leonel1}.

The paper is organized as follows. In Section \ref{sec2} we discuss the mapping 
and the properties of the phase space. Some properties of the Lyapunov exponent 
as well as the Lyapunov time are discussed in Section \ref{sec3}. The 
Poincar\'e recurrence time is discussed in Section \ref{sec4} while discussions 
and final remarks are drawn in Section \ref{sec5}.

\section{The model, the mapping and the phase space}
\label{sec2}

The Fermi-Ulam model is composed of a classical particle confined to move 
inside of two rigid walls. One is considered fixed at $x=\ell$ while the other 
is periodically moving whose position is given by $x_w(t)=X_0\cos(\omega 
t)$ where $X_0$ is the amplitude of the motion and $\omega$ is the frequency of 
oscillation. The particle experiences elastic collisions with the wall. The 
dynamics of the particle is given by a two dimensional, nonlinear and area 
preserving mapping describing how the velocity of the particle and phase of the 
moving wall transform from the impact $n$ to the impact $n+1$. The version 
of the model we consider in this paper is the so called static wall 
approximation \cite{karlis}. It assumes that, because of the small range of 
values considered for the control parameter $X_0$, both walls are 
considered fixed. However when a particle collides with one wall at the left 
it suffers an exchange of energy and momentum as if the wall were moving. This 
version of the model retains the majority of the properties of the whole 
version where the moving wall is taking into account, including localization of 
the periodic regions, determination of the position of the invariant spanning 
curves and the scaling produced by it \cite{leonel}. However there is a huge 
advantage of speeding up the numerical simulations the static wall 
approximation has as compared to the complete model where transcendental 
equations are compulsory to be solved.

Considering a set of dimensionless variables such as $\varepsilon=X_0/\ell$, 
$V_n=v_n/(\omega\ell)$ with $v_n$ representing the velocity of the particle and 
$\phi=\omega t$ the mapping describing the dynamics of the model is written as
\begin{equation}\label{modelo}
    T: \left\{
  \begin{array}{c}
     \phi_{n+1}= \left[\phi_{n}+\frac{2}{V_n}\right] {\rm mod}~~2\pi \\
     V_{n+1}= \left| V_n -2\varepsilon \sin(\phi_{n+1}) \right| 
  \end{array}
  \right.,
\end{equation}
where the absolute value in the second equation was introduced as an attempt to 
avoid that, after a collision, a particle has negative velocity 
\cite{leonel2019}.

\begin{figure}[t]
\begin{center}
\includegraphics[width=1.0\linewidth]{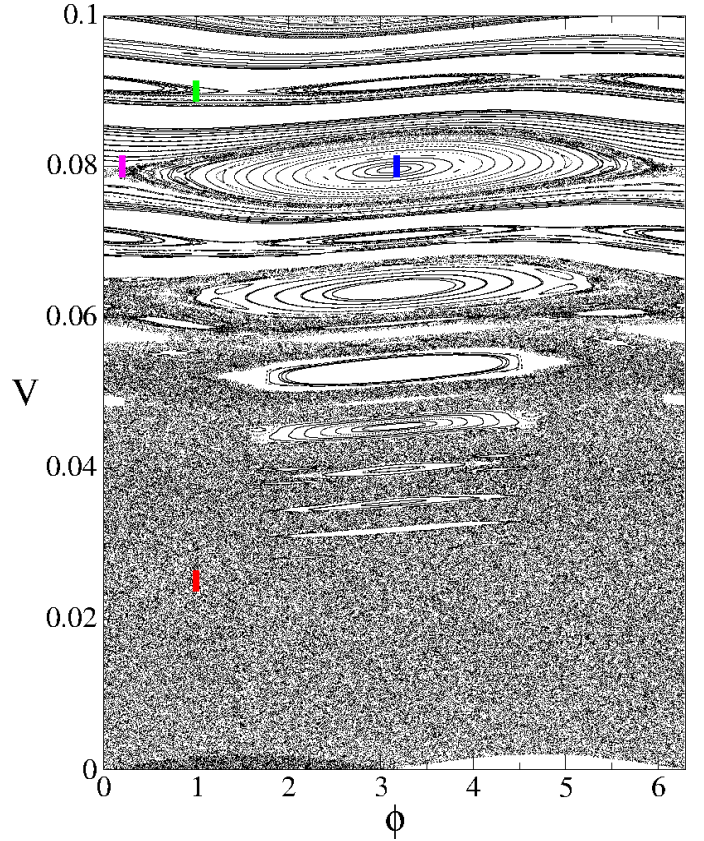}
\caption{\label{fase}Plot of the phase space for the static wall 
approximation of the Fermi-Ulam Model. The control parameter used was 
$\varepsilon=10^{-3}$.}
\end{center}
\end{figure}

The phase space of the model is shown in Figure \ref{fase} for the control 
parameter $\varepsilon=10^{-3}$. It is easy to note a mixed structure of it 
including the presence of a large chaotic region coexisting with periodic 
structures such as elliptical islands and also invariant spanning curves. There 
are four regions in the phase space identified in the figure corresponding to 
the domains we are considering in the investigation along this paper. The size 
of the chaotic sea is marked by the minimal region as the lowest velocity of 
the 
wall while the upper limit is determined by the smallest velocity energy 
invariant spanning curve. Above the curve one observe local chaos while below 
of it there is global chaos. According to the Chirikov criteria 
\cite{lichtenberg1992}, the 
last invariant spanning curve broken in the so called Standard Mapping 
\cite{lichtenberg1992} happens at a critical parameter $K_c\cong 0.9716\ldots$. 
The standard 
mapping is written as $I_{n+1}=I_n+K\sin(\theta_{n+1})$ and 
$\theta_{n+1}=(\theta_n+I_{n})~{\rm mod}~2\pi$. When the second equation of the 
mapping (\ref{modelo}) is written replacing $V_n=V^*+\Delta V_n$ with $(\Delta 
V)/V^*\ll 1$ and Taylor expanding it till first order and that when compared 
with the equations of the standard mapping leads to 
$V^*={{2}\over{\sqrt{0.9716\ldots}}}\sqrt{\varepsilon}$. The exponent heading 
$\varepsilon$ plays a major rule on the regime of growth and saturation of the 
curves for the average velocity. As discussed in Ref. \cite{leonel}, the 
exponent of the curves of $\bar{V}_{\rm sat}\propto \varepsilon^{\alpha}$ with 
$\alpha=1/2$ which is one of the three critical exponents. The exponent marking 
the diffusion for low velocity is $V\propto(n\epsilon^2)^{\beta}$ with 
$\beta=1/2$. The last exponent is obtained by a scaling law $z=\alpha/\beta-2$.

In the next section we discuss the characteristic Lyapunov time based on the 
results for the positive Lyapunov exponent.

%***********************************************************
\section{The Lyapunov Time}
\label{sec3}

In this section we discuss our results for the characteristic Lyapunov time, 
which is defined as the inverse of the positive Lyapunov exponent. Indeed the 
Lyapunov exponent is a common measure to estimate how chaotic a system is. A 
positive Lyapunov exponent yields in an exponentially fast spread of two very 
near initial conditions in the phase space. For a two dimensional mapping 
they can be obtained \cite{eckmann} using the eigenvalues of the Jacobian matrix
\begin{equation}
\lambda_j=\lim_{n\rightarrow\infty}{{1}\over{n}}\ln|\Lambda_n^{(j)}|,
\label{c10_eq20}
\end{equation}
with $j=1,2$ where $\Lambda_n^{(j)}$ correspond to the eigenvalue of 
the Jacobian matrix $M=\Pi_{i=1}^nJ_i(V_i,\phi_i)=J_nJ_{n-1}J_{n-2}\ldots 
J_2J_1$. Since the convergence of the Lypaunov exponent is observed for large 
$n$ the accumulation of the product of the $J_i$ matrices may lead to 
overflow in their coefficients hence making hard the estimation of $\lambda$. 
The triangularization algorithm avoid such a trouble. It consists of rewrite 
$J$ as $J=\Theta T$ with $\Theta$ being an orthogonal matrix obeying the 
property of $\Theta^{-1}=\Theta^T$ and $T$ is a triangular matrix. Therefore 
this leads to
$$
\Theta=\left(\begin{array}{ll}
\cos(\theta)  &  -\sin(\theta)  \\
\sin(\theta)  &    \cos(\theta)\\
\end{array}
\right),
$$
with
$$
T=\left(\begin{array}{ll}
T_{11}  &  T_{12}  \\
0  &    T_{22}\\
\end{array}
\right).
$$

We notice the matrix $M$ can be written as
\begin{eqnarray}
M&=&J_nJ_{n-1}J_{n-2}\ldots J_2J_1,\nonumber\\
&=&J_nJ_{n-1}J_{n-2}\ldots J_2\Theta_1\Theta_1^{-1}J_1.
\end{eqnarray}
Defining $T_1=\Theta_1^{-1}J_1$ and $\tilde{J}_2=J_2\Theta_1$ the coefficients 
of $T_1$ are
$$
\left(\begin{array}{ll}
T_{11}  &  T_{12}  \\
0  &    T_{22}\\
\end{array}
\right)=
\left(\begin{array}{ll}
\cos(\theta)  &  \sin(\theta)  \\
-\sin(\theta)  &   \cos(\theta)\\
\end{array}
\right)
\left(\begin{array}{ll}
j_{11}  &  j_{12}  \\
j_{21}  &    j_{22}\\
\end{array}
\right).
$$
From $T_{21}=0$ we end up with $0=-j_{11}\sin(\theta)+j_{21}\cos(\theta)$ 
yielding in
\begin{equation}
{{j_{21}}\over{j_{11}}}={{\sin(\theta)}\over{\cos(\theta)}}.
\label{c10_eq21}
\end{equation}
Instead of using $\theta=\arctan(j_{21}/j_{11})$ which is rather expensive 
numerical function, we use the expressions of $\sin(\theta)$ and $\cos(\theta)$ 
directly from $J$, hence
\begin{eqnarray}
\cos(\theta)&=&{{j_{11}}\over{\sqrt{j_{11}^2+j_{21}^2}}},\\
\sin(\theta)&=&{{j_{21}}\over{\sqrt{j_{11}^2+j_{21}^2}}}.
\end{eqnarray}
The expressions for $T_{11}$ and $T_{22}$ can be written as 
$T_{11}=j_{11}\cos(\theta)+j_{21}\sin(\theta)$ and also
$T_{22}=-j_{12}\sin(\theta)+j_{22}\cos(\theta)$ producing the following 
expressions
\begin{eqnarray}
T_{11}&=&{{j_{11}^2+j_{21}^2}\over{\sqrt{j_{11}^2+j_{21}^2}}},\\
T_{22}&=&{{j_{11}j_{22}-j_{12}j_{21}}\over{\sqrt{j_{11}^2+j_{21}^2}}}.
\end{eqnarray}
Once $T_{11}$ and $T_{22}$ are known the matrix $\tilde{J}_2$ is given by
$\tilde{J}_2=J_2\Theta_1$
$$
\left(\begin{array}{ll}
\tilde{j}_{11}  &  \tilde{j}_{12}  \\
\tilde{j}_{21}  &  \tilde{j}_{22}\\
\end{array}
\right)=
\left(\begin{array}{ll}
j_{11}  &  j_{12}  \\
j_{21}  &  j_{22}\\
\end{array}
\right)
\left(\begin{array}{ll}
\cos(\theta)  &  -\sin(\theta)  \\
\sin(\theta)  &  \cos(\theta)\\
\end{array}
\right).
$$
The procedure is then repeated for the second iteration, and third and any 
further iteration of the mapping until the complete series of matrices is 
exhausted. The Lyapunov exponents are then given by
\begin{equation}
\lambda_{j}=\lim_{n\rightarrow\infty}\sum_{i=1}^n\ln|T_{jj}^{(i)}|,~j=1,
2.
\end{equation}

\begin{figure}[t]
\begin{center}
\includegraphics[width=1.0\linewidth]{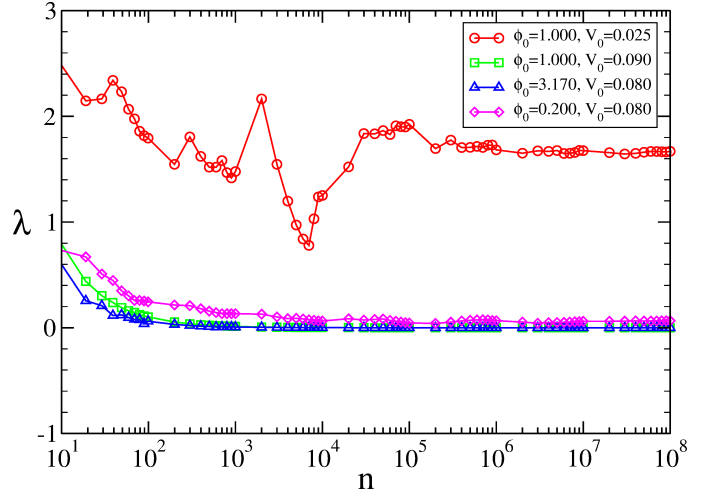}
\caption{\label{lyap} Plot of the time evolution of the positive Lyapunov 
exponent for the regions marked in the phase space of figure \ref{fase}. The 
control parameter used was $\varepsilon=10^{-3}$. Each initial condition was 
evolved up to $n=10^8$ collisions with the walls.}
\end{center}
\end{figure}
Figure \ref{lyap} shows the converge of the positive Lyapunov exponent using 
the algorithm discussed above for the specific regions defined in the phase 
space. Region 1 is marked as the red color in the Figure and is evolved from 
the 
initial condition $(\phi_0=1.000, V_0=0.025)$, while region 2 marked as green 
is for $(\phi_0=1.000, V_0=0.090)$, blue is for region 3 with $(\phi_0=3.170, 
V_0=0.080)$ and finally magenta is for region 4 obtained for $(\phi_0=0.200, 
V_0=0.080)$. A chaotic domain leads to a convergence of 
$\overline{\lambda}=1.665(3)$ while periodic regions lead to Lyapunov exponent 
converging to null value. Table \ref{tabela} summarizes both the Lyapunov 
exponent as well as the Lyapunov time which is defined as the inverse of the 
Lyapunov exponent $t_L=1/\lambda$.  It is important to notice that for the 
chaotic regions ($R_1$ and $R_4$) the Lyapunov time is relatively short, so 
returning to Lyapunov time concept it is clear that for these regions the map 
quickly shows a chaotic behavior. However for the regions near the islands of 
stability ($R_2$ and $R_3$) the Lyapunov time is significantly large indicating 
absence of chaos. We emphasize both $\lambda$ and $t_L$ were calculated for a 
finite number of iterations of $10^8$. 

\begin{figure*}
\subfigure[\label{fase1}]{
\includegraphics[width=0.45\linewidth]{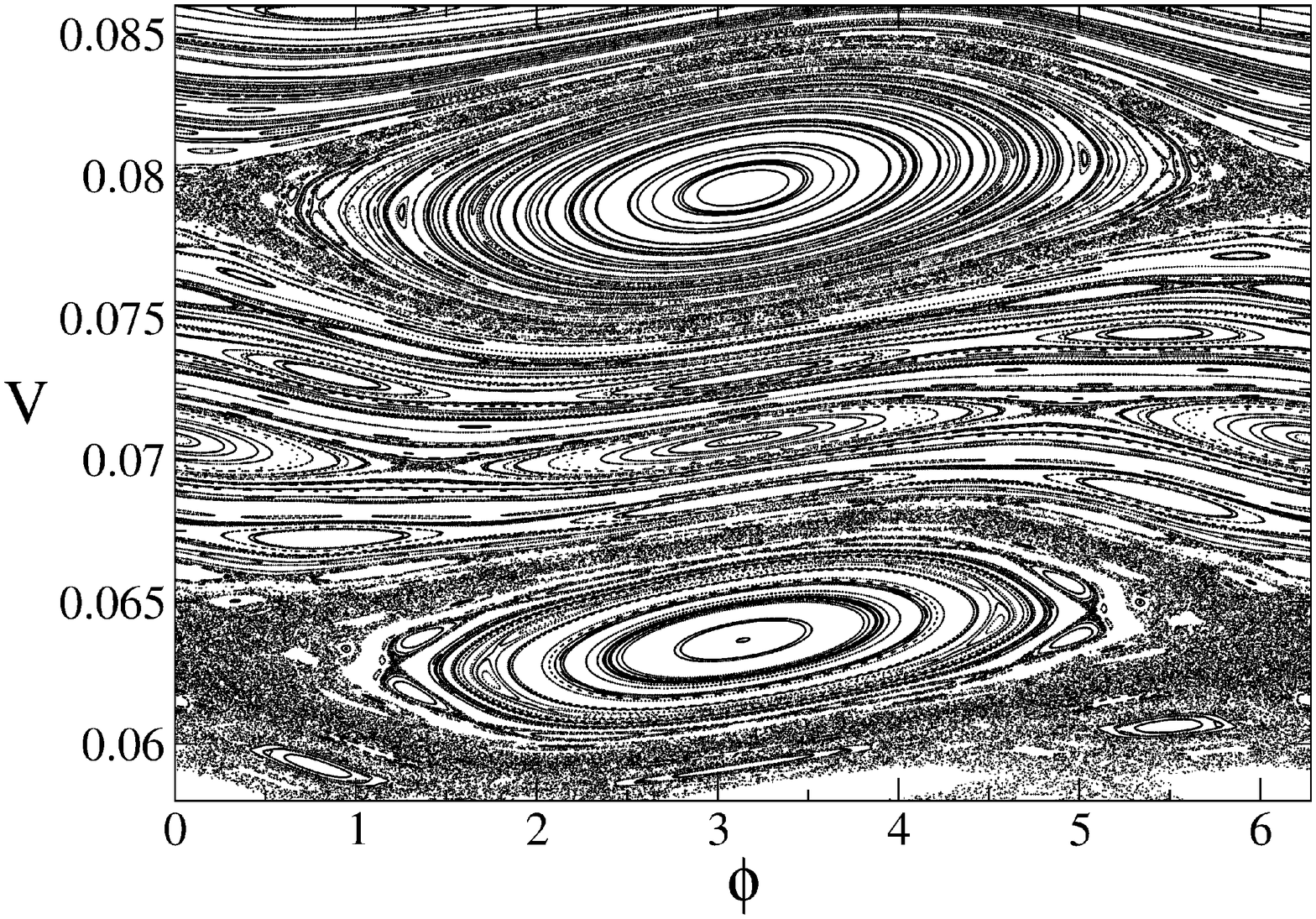}}
\subfigure[\label{fase2}]{
\includegraphics[width=0.52\linewidth]{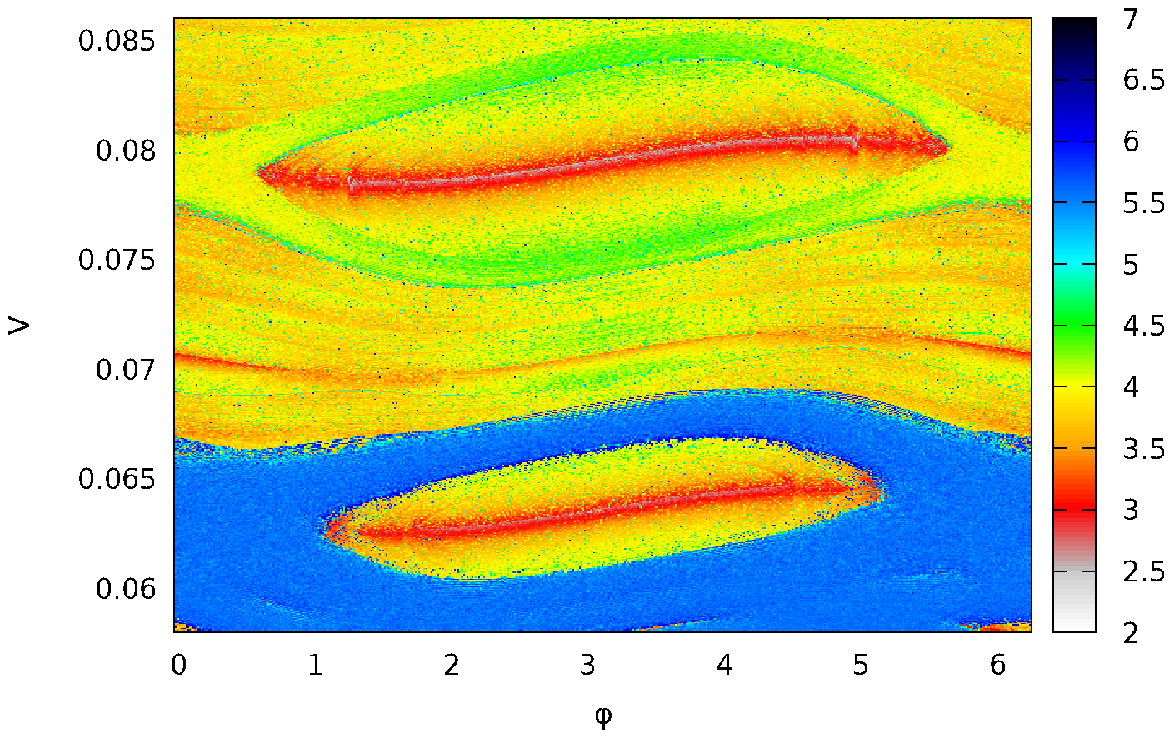}}
\caption{(a) Plot of an expanded region of the phase space shown in Fig. 
\ref{fase} where the two largest islands are observed, one below and other 
above the first invariant spanning curve. (b) Plot of the same region of 
(a) with the color scale representing in logarithmic scale the mean Poincar\'e 
recurrence time $\langle t_r \rangle$.}
\end{figure*}

\begin{table}[h]
\begin{center}
\caption{\label{tabela} Numerical values of $\lambda$ and $t_L$ for the regions 
indicated in 
the phase space of the figure \ref{fase}.}
\begin{tabular}{@{}cll@{}}
\toprule
Regions & \multicolumn{1}{c}{$\lambda$}   & \multicolumn{1}{c}{$t_L=1/\lambda$} 
\\ \midrule
$R_1$    & $ 1.665(3)$ &   $ 0.600(1)$   \\ 
$R_2$    & $ 1.598(1) \times 10^{-6}$ &  $62(5) \times 10^{5}$   \\
$R_3$    & $ 4.258(3) \times 10^{-7}$ &  $2348(4) \times 10^{3}$    \\
$R_4$    & $ 6.627(4) \times 10^{-2}$ &  $15.089(9)$ \\ \bottomrule
\end{tabular}
\end{center}

\end{table}

\begin{figure}[t]
\begin{center}
\includegraphics[width=1.0\linewidth]{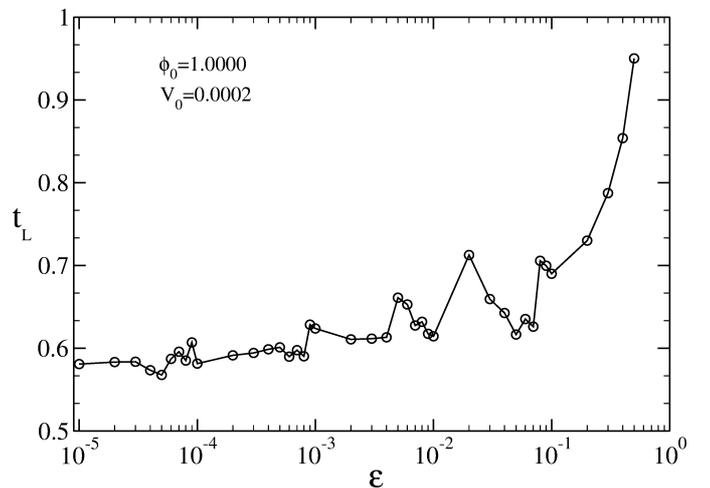}
\caption{\label{tlyap} Plot of the numerical value of $t_L$ as a function of 
$\varepsilon$ measured for a chaotic region of the phase space.}
\end{center}
\end{figure}

Figure \ref{tlyap} shows a plot of $t_L~vs.~\varepsilon$. Each point of the 
curve was obtained after a long simulation of $10^8$ iterations for the 
calculation of $\lambda$. One sees that $t_L$ increases, in the average, with 
the increase of $\varepsilon$. The regime of growth for $t_L$ is slow at the 
beginning and speeds up for $\varepsilon>10^{-1}$. The latter window of 
control parameter the static wall approximation has severe limitations since 
the movement of the time dependent wall would indeed affects the shape of the 
phase space leading to frequent non physical situations.

%************************************************************
\section{The Poincar\'e Recurrence Time}
\label{sec4}

In this section we discuss the Poincar\'e recurrence time. The essence is 
simple and consists in determine the time a particle which left a given region 
of the phase space returns to a point $\delta$-close to that region. An orbit 
in the phase space is said to recur to an $I_{\delta}$ interval if, once it 
starts at point $\overrightarrow{x}_0 \in I_{\delta}$, $\forall \delta$ 
with $\overrightarrow{x}_0=(V_0,\phi_0)$ there is a time $t^*$ such that, after 
$t^*$, the orbit is at a distance $|\overrightarrow{x}_{t^*} - 
\overrightarrow{x}_0|\leq \delta$, hence $\overrightarrow{x}_{t^*} \in 
I_{\delta}$ (see Ref. \cite{altmann2}). Figure \ref{fase1} 
shows an expanded domain of the phase space plotted in Figure \ref{fase} where 
two period-1 islands are present being one below and another above of the first 
invariant spanning curve. We notice also that in between them there is a chain 
of smaller islands and some chaotic regions around them hence 
characteristic of a mixed phase space. Figure \ref{fase2} plots the same region 
of Figure \ref{fase1} but with the color scheme defined as the mean Poincar\'e 
recurrence time $\langle t_r \rangle$ plotted in logarithmic scale. From figure 
\ref{fase2} it is possible to notice a separation of two regions of the phase 
space, one in blue (dark gray) indicating that $\langle t_r \rangle$ is between 
$10^5$ and $10^6$ while in the second, in orange (light gray), giving $\langle 
t_r \rangle$ between $10^3$ and $10^4$ iterations. It is worth mentioning that 
the stickiness phenomenon can affect the recurrence time. This is because the 
orbit stays stuck in the certain region of the phase space until it escapes 
such domain and eventually returns to a position close to the initial 
condition. This can be confirmed looking at figure \ref{fase2} and seeing that 
near the islands where stickiness is observed the recurrence time is longer as 
compared to other regions and clearly identified in the color scale.

Other interesting physical measure from the Poincar\'e recurrence time is 
linked to the fractal dimension of the region since a power law fitting of 
$\langle t_r \rangle ~vs. ~\epsilon$ gives an exponent which is the absolute 
value of the fractal dimension $d_w$ whenever the limit of $\epsilon\rightarrow 
0$ is considered \cite{Harsoula2018}. The parameter $\epsilon$ corresponds to 
the size of the recurrent window in the phase space. The relationship between 
$t_r$ and the chosen region is given as
\begin{equation}\label{dimensao}
    \langle t_r \rangle=\frac{1}{\epsilon^{d_w}},
\end{equation}
where $d_w$ is the fractal dimension and $\epsilon$ the side of the box 
\cite{Harsoula2018}.

\begin{figure}
\begin{center}
\includegraphics[width=1.0\linewidth]{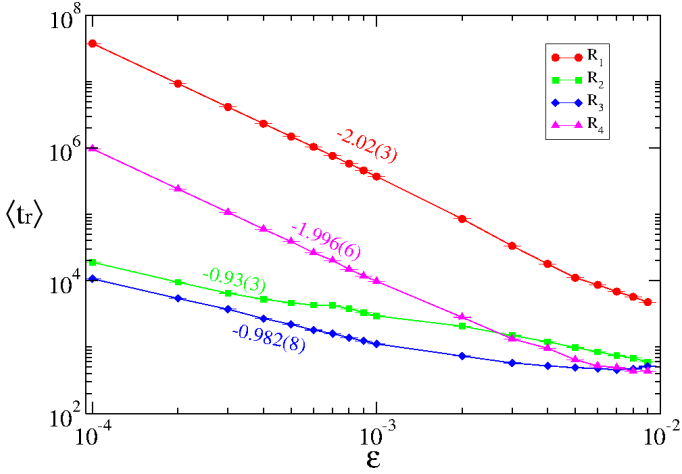}
\caption{\label{tr} Plot of the average Poincar\'e recurrence time $\langle t_r 
\rangle$ as a function of the size of the recurrence interval $\epsilon$, for 
the different regions indicated in figure \ref{fase}.}
\end{center}
\end{figure}

It is shown in figure \ref{tr} a plot of $\langle t_r \rangle~vs.~\epsilon$ 
where two different slopes are observed. For the curves related to the initial 
conditions given in the chaotic regions a power law fitting gives a decay with  
slope $\approx-2$ while for the initial conditions given in the regions on the 
periodic islands the slope of the decay is $\approx-1$. The fractal dimension is 
related to the diffusion coefficient $\mu$ through the following equation 
\cite{Harsoula2018}
\begin{equation}\label{difusao}
d_w=\frac{2}{\mu}.
\end{equation}
When the initial conditions are located along the islands of 
periodicity, $d_w=1$ yielding $\mu=2$ which is a signature of anomalous 
diffusion. On the other hand for initial conditions placed in the chaotic 
region, $d_w=2$ leading to $\mu=1$ and one observed the chaotic orbits 
experience normal diffusion.

%***********************************************************
\section{Discussion and Final remarks}
\label{sec5}

Let us now discuss the implications of the results obtained. Unlike to what 
happens for the standard map where the Lyapunov exponent $\lambda$ increases 
as the nonlinearity increases \cite{Harsoula2018}, in the Fermi-Ulam model the 
Lyapunov exponent decreases and consequently, the Lyapunov $t_L$ time increases 
with the increase of the nonlinearity. This is due to the fact that in the 
standard map the increase in the nonlinearity parameter causes the invariant 
curves and stable structures to be destroyed and chaos dominate over the 
system, 
whereas in the Fermi-Ulam model, the increase in the nonlinearity does 
not break the invariant spanning curve. Chaotic sea is scaling with the 
localization of the first invariant spanning curve hence of the type 
$\sqrt{\varepsilon}$.

Regarding the Poincar\'e recurrence time $t_r$ the results found by 
\cite{Harsoula2018} for the standard map were also observed in the Fermi-Uam 
model confirming $\langle t_r \rangle$ scales with $\epsilon$ which is the 
size of the recurrence domain. For chaotic regions the slope obtained is about 
$\mu=2$ while for periodic regions is $\mu=1$.

As a final remark, we have measured both the Lyapunov $t_L$ time as the inverse 
of the Lyapunov exponent for different regions of the phase space. The average  
Poincar\'e recurrence was confirmed to be dependent on the size of the 
recurrent window and where the initial condition was given. A power law fitting 
of $\langle t_r \rangle~vs.~\epsilon$ gives a slope of $-2$ for initial 
conditions taken in chaotic region while yields a slope of $-1$ for 
periodic domains. We saw that $d_w=\frac{2}{\mu}$, so that in the chaotic 
regions the diffusion exponent converges to $\mu=1$ and within the stability 
islands $\mu=2$.

%*******************************************************

\section*{ACKNOWLEDGMENTS}

Instituto Federal de Educa\c c\~ao, Ci\^encia e Tecnologia do Sul de Minas 
Gerais, IFSULDEMINAS - Campus Inconfidentes. Grupo de Escudos em Modelagem 
Computacional e Aplica\c c\~oes - GEMCA. EDL thanks to CNPq (301318/2019-0) and 
FAPESP (2019/14038-6), Brazilian agencies.

%*******************************************************
\bibliographystyle{unsrt}
\bibliography{ref}

\providecommand{\noopsort}[1]{}\providecommand{\singleletter}[1]{#1}%
\begin{thebibliography}{10}

\bibitem{p2}
Kenya Murase, Shuji Tanada, Hiroshi Mogami, Masashi Kawamura, Masao Miyagawa,
  Masafumi Yamada, Hiroshi Higashino, Atsushi Iio, and Ken Hamamoto.
\newblock Validity of microsphere model in cerebral blood flow measurement
  using n-isopropyl-p-(i-123) iodoamphetamine.
\newblock {\em Medical physics}, 17(1):79--83, 1990.

\bibitem{p5}
C~Hagedorn, EL~Mc~Coy, and TM~Rahe.
\newblock The potential for ground water contamination from septic effluents 1.
\newblock {\em Journal of Environmental Quality}, 10(1):1--8, 1981.

\bibitem{p3}
William~F Morris.
\newblock Predicting the consequence of plant spacing and biased movement for
  pollen dispersal by honey bees.
\newblock {\em Ecology}, 74(2):493--500, 1993.

\bibitem{p4}
David Popp.
\newblock International innovation and diffusion of air pollution control
  technologies: the effects of nox and so2 regulation in the us, japan, and
  germany.
\newblock {\em Journal of Environmental Economics and Management},
  51(1):46--71, 2006.

\bibitem{b1}
Rostislav~Vsevolodovich Ozmidov.
\newblock Diffusion of contaminants in the ocean.
\newblock 1990.

\bibitem{b2}
Venkataraman Balakrishnan.
\newblock {\em Elements of nonequilibrium statistical mechanics}, volume~3.
\newblock Ane Books, 2008.

\bibitem{celia}
Edson D. Leonel and Célia M. Kuwana.
\newblock An investigation of chaotic diffusion in a family of hamiltonian
  mappings whose angles diverge in the limit of vanishingly action.
\newblock {\em Journal of Statistical Physics}, 170:69--78, 2018.

\bibitem{lichtenberg1992}
A.J. Lichtenberg and M.A. Lieberman.
\newblock {\em Regular and chaotic dynamics}.
\newblock Applied mathematical sciences. Springer-Verlag, 1992.

\bibitem{Harsoula2018a}
Mirella Harsoula and George Contopoulos.
\newblock Global and local diffusion in the standard map.
\newblock {\em Physical Review E}, 97(2):022215, 2018.

\bibitem{altmann1}
Eduardo~G Altmann, Adilson~E Motter, and Holger Kantz.
\newblock Stickiness in mushroom billiards.
\newblock {\em Chaos: An Interdisciplinary Journal of Nonlinear Science},
  15(3):033105, 2005.

\bibitem{buni1}
Leonid~A Bunimovich.
\newblock Fine structure of sticky sets in mushroom billiards.
\newblock {\em Journal of Statistical Physics}, 154(1-2):421--431, 2014.

\bibitem{zas1}
George~M Zaslavsky.
\newblock {\em The physics of chaos in Hamiltonian systems}.
\newblock world scientific, 2007.

\bibitem{chaos_b}
Arkady Pikovsky and Antonio Politi.
\newblock {\em Lyapunov exponents: a tool to explore complex dynamics}.
\newblock Cambridge University Press, 2016.

\bibitem{contopoulous_2}
Henry~E. Kandrup, Christos Siopis, G.~Contopoulos, and Rudolf Dvorak.
\newblock Diffusion and scaling in escapes from two-degrees-of-freedom
  hamiltonian systems.
\newblock {\em Chaos: An Interdisciplinary Journal of Nonlinear Science},
  9(2):381--392, 1999.

\bibitem{liouville}
F.~Reif.
\newblock {\em Fundamentals of statistical and thermal physics / [by] F. Reif}.
\newblock McGraw-Hill Kogakusha Tokyo, international student ed. edition, 1965.

\bibitem{fermi}
ENRICO Fermi.
\newblock On the origin of the cosmic radiation.
\newblock {\em Phys. Rev.}, 75:1169--1174, Apr 1949.

\bibitem{leonel}
E.~D. Leonel and P.~V.~E. McClintock.
\newblock Chaotic properties of a time-modulated barrier.
\newblock {\em Phys. Rev. E}, 70:016214, Jul 2004.

\bibitem{leonel1}
Edson~D Leonel, Juliano~A de~Oliveira, and Farhan Saif.
\newblock Critical exponents for a transition from integrability to
  non-integrability via localization of invariant tori in the hamiltonian
  system.
\newblock {\em Journal of Physics A: Mathematical and Theoretical},
  44(30):302001, jul 2011.

\bibitem{karlis}
A.~K. Karlis, P.~K. Papachristou, F.~K. Diakonos, V.~Constantoudis, and
  P.~Schmelcher.
\newblock Hyperacceleration in a stochastic fermi-ulam model.
\newblock {\em Phys. Rev. Lett.}, 97:194102, Nov 2006.

\bibitem{leonel2019}
Edson~Denis Leonel.
\newblock {\em Invariância de Escala em Sistemas Dinâmicos Não Lineares}.
\newblock Blucher, first edition, 2019.

\bibitem{eckmann}
J.~P. Eckmann and D.~Ruelle.
\newblock Ergodic theory of chaos and strange attractors.
\newblock {\em Rev. Mod. Phys.}, 57:617--656, Jul 1985.

\bibitem{altmann2}
Eduardo~G. Altmann, Elton~C. da~Silva, and Iberê~L. Caldas.
\newblock Recurrence time statistics for finite size intervals.
\newblock {\em Chaos: An Interdisciplinary Journal of Nonlinear Science},
  14(4):975--981, 2004.

\bibitem{Harsoula2018}
Mirella Harsoula, Kostas Karamanos, and George Contopoulos.
\newblock Characteristic times in the standard map.
\newblock {\em Physical Review E}, 99(3):032203, 2019.

\end{thebibliography}
%********************************************************

\end{document}